\documentclass{elsart}
\usepackage{graphicx,natbib,amssymb}
\journal{New Astronomy}
\def\astrobj#1{#1}
\begin{document}
\begin{frontmatter}

\title{An Investigation of 11 previously unstudied open star clusters}

\author{Tadross, A. L.\thanksref{fn1}}

\thanks[fn1]{Email: altadross@nriag.sci.eg --- altadross@yahoo.com \\
             Phone: +202 25560645 --- Fax: +202 25548020}

\thanks[fn2]{doi:xxxx/j.newast.2008.xx.xxx}

\address{National Research Institute of Astronomy and Geophysics,
11421 - Helwan, Cairo, Egypt. \\[2mm]
    {\rm Received xx xxxxx 2008; accepted xx xxxxx 2008}\thanksref{fn2}}

\begin{abstract}
The main astrophysical properties of 11 previously unstudied open
star clusters are probed with {\it JHK} Near-IR {\it (2MASS)}
photometry of Cutri et al. [Cutri, R., et al., 2003. The IRSA 2MASS All-sky Point Source
Catalog, NASA/IPAC Infrared Science Archive] and proper motions (NOMAD) astrometry of Zacharias et al. [Zacharias, N., Monet, D., Levine, S., Urban, S., Gaume, R., Wycoff, G., 2004. American Astro. Soc. Meeting 36, 1418]. The fundamental parameters have been derived for \astrobj{IC (1434, 2156)}; \astrobj{King (17, 18, 20, 23, 26)}; and \astrobj{Dias (2, 3, 4, 7, 8)}, for which no prior parameters are available in the literature. The clusters' centers coordinates and angular diameters are re-determined, while ages, distances, and color excesses for these clusters are estimated here for the first time.
\end{abstract}

\begin{keyword}
Galaxy: open clusters and associations: general -- individual: ID
clusters (IC - King - Dias) -- Stars: Hertzsprung-Russell (HR)
diagram

\PACS 91.10.Lh \sep 95.80.+p \sep 95.85.Jq \sep 97.10.Zr \sep
98.20.Di
\end{keyword}
\end{frontmatter}

\section{Introduction}
According to some estimations, there are as many as 100,000 open
star clusters in the Galaxy, but less than 2000 of them have been
discovered and cataloged, Glushkova et al. (2007). Actually, not
all the discovered clusters have their basic photometrical
parameters in the current literatures indeed. So, our aim in the
previous and present continuation series of papers to determine
the main astrophysical properties of rarely or unstudied open
star clusters using modern databases (cf. Tadross 2008 and
references therein; hereafter TA08). On this respect, the present
study introduces the first photometric analysis of the color
magnitude diagrams {\it (CMDs)} of the clusters under
investigation.
\\ \\
The Naval Observatory Merged Astrometric Dataset {\it (NOMAD)} of Zacharias et al. (2004) and the Two Micron All Sky Survey {\it (2MASS)} of Cutri et al. (2003) are used to determine the fundamental parameters of 11 open star clusters of IC, King, and
Dias; which were never studied insofar. The only information known about these clusters are their centers' coordinates and sometimes their apparent diameters; listed here in Table 1; which are provided by Mermilliod (1995) and Dias (2002) catalogs. These catalogs are
constantly updated and maintained in electronic form (WEBDA\footnote{\it http://obswww.unige.ch/webda} and DIAS\footnote{\it
http://www.astro.iag.usp.br/$^\sim$wilton/clusters.txt} sites).
Note that cluster King 20 has no parameters in WEBDA, but only in Dias catalog. Hence, we used this cluster to calibrate our reductions' procedure. For that purpose, the basic parameters of King 20 are re-estimated and compared with those available in the literature (Bica et al. 2006). The derived parameters are found very close to the published ones, which make our reductions' procedure is fairly acceptable.

\section{Data Reductions Procedure}

Data extraction have been performed using the known tool of VizieR
Catalogs of (2MASS)\footnote{\it
http://vizier.u-strasbg.fr/viz-bin/VizieR?-source=2MASS} and
(NOMAD)\footnote{\it
http://vizier.cfa.harvard.edu/viz-bin/VizieR?-source=I/297/}. The
investigated clusters have been selected from WEBDA and
DIAS catalogs under the following conditions:\\
(1) The clusters' data have been extracted at a preliminary radii
of about 10 arcmin from their obtained centers; \\
(2) The clusters should have good blue images on Digitized Sky
Surveys {\it DSS}\,\footnote {\it
http://cadcwww.dao.nrc.ca/cadcbin/getdss} and clearly
distinguished from the background field, (see Fig. 1 in TA08);\\
(3) The foreground stars have been separated from the cluster stars using the proper motion data of {\it (NOMAD)} dataset (i.e., all stars with nonzero proper
motions and those distributed over the field with no concentration around
the cluster center have been removed);\\
(4) The clusters should have enough members with prominent
sequences in their {\it CMDs};\\
(5) A cutoff of photometric completeness limit at $J<16.5$ mag is
applied on the {\it (2MASS)} data to avoid the over-sampling (cf.
Bonatto et al. 2004);\\
(6) The stars with observational errors more than 0.20\, mag are cancelled; and\\
(7) According to Bonatto et al. (2005), the membership criteria is
adopted for the location of the stars in the {\it CMDs}.
Therefore, color-magnitude filters have been applied to the
$J\sim(J-H)$ and $K\sim(J-K)$ sequences to isolate probable member
stars, whereas the stars located away from the main sequences are
excluded. The maximum departure accepted here is about 0.15 mag.

\section{Astrometry}
To derive better cluster center, the cluster center is
re-determined and taken at the maximum stellar density of the
cluster's area. The location of the cluster center is found by
fitting a Gaussian to the profiles of star counts in right
ascension $\alpha$ and declination $\delta$, (see Fig. 3 in TA08).
\\

Within concentric shells in equal incremental steps from the
cluster center, the radial stellar density distribution is
performed out to the preliminary radius. Density distributions of
all our sample are well-represented by King (1962) profiles. The
real radius (genuine border) of the cluster can be defined at that
point which reaches enough stability of the background density and
covers all the cluster area. At that radius, the {\it JHK}
photometric data would be extracted and taken into account for
estimating the photometrical clusters' properties, see Fig. 4 in
TA08. The estimated clusters centers and diameters are shown in Table
2; columns 2-3 and 4 respectively.

\section{Photometry}
Most of the clusters' sample have lower galactic latitudes,
whereas the background field is found very crowded and
consequently {\it CMDs}~ are found very contaminated as well.
Therefore, the {\it CMDs} profiles are comprising stars inside
radii of 1$^{'}$, 2$^{'}$, and 3$^{'}$ from the cluster center.
The simultaneous fitting of the solar metallicity isochrones of
Bonatto et al. (2004) are attempted on $J\sim(J-H)$ and
$K\sim(J-K)$ diagrams for the inner stars, in which they should be
less contaminated by field stars. If the number of stars are not
enough for an accepted fitting, the next larger area would be
included, and so on. Moreover, the reddening estimation values have
been guided by the galactic absorption values of Schlegel et al. (1998).
\\ \\
R$_{V}$= 3.2, $\frac {A_{J}}{A_{V}}$= 0.276, $\frac
{A_{K}}{A_{V}}$= 0.118 and $\frac {E_{J-H}}{E_{B-V}}$= 0.33 have
been used for reddening and absorption transformations according
to Dutra et al. (2002) and references therein. $\frac
{E_{J-K}}{E_{J-H}}\approx$ 1.6 $\pm$ 0.10, which was derived from
absorption rations in Schlegel et al. (1998).

\section{Conclusion}
Determining the clusters main parameters with the present reductions showing a good agreement with that published for the previously studied cluster \astrobj{King 20}. Following the above procedure, the cluster center, diameter, age, reddening, distance modulus, distance from the sun, distance from the galactic center, $R_{gc}$, and the
projected distances on the galactic plane from the Sun, $X_{\odot}$, $Y_{\odot}$, and the distance from galactic plane, $Z_{\odot}$ of all our clusters' have been estimated and listed here in Table 2. The {\it CMDs} and isochrone fits for all the investigated clusters can be seen in Figs. 1-3.


\section*{Acknowledgements}

This manuscript is presented as a poster article without details
in the {\it First Middle East-Africa, Regional IAU Meeting, Cairo,
Egypt, April 5-10, 2008} (arXiv:0804.2567). This publication makes
use of data products from the Naval Observatory Merged Astrometric
Dataset {\it NOMAD} of Zacharias et al. (2004), and the Two Micron
All Sky Survey {\it 2MASS} of Cutri et al. (2003), which is a
joint project of the University of Massachusetts and the Infrared
Processing and Analysis Center/California Institute of Technology,
funded by the National Aeronautics and Space Administration and
the National Science Foundation. Catalogues from {\it CDS}/{\it
SIMBAD} (Strasbourg), and Digitized Sky Survey {\it DSS} images
from the Space Telescope Science Institute have been employed.

\newpage
   \begin{figure*}
\begin{center}
      {\includegraphics[width=15cm,height=22cm]{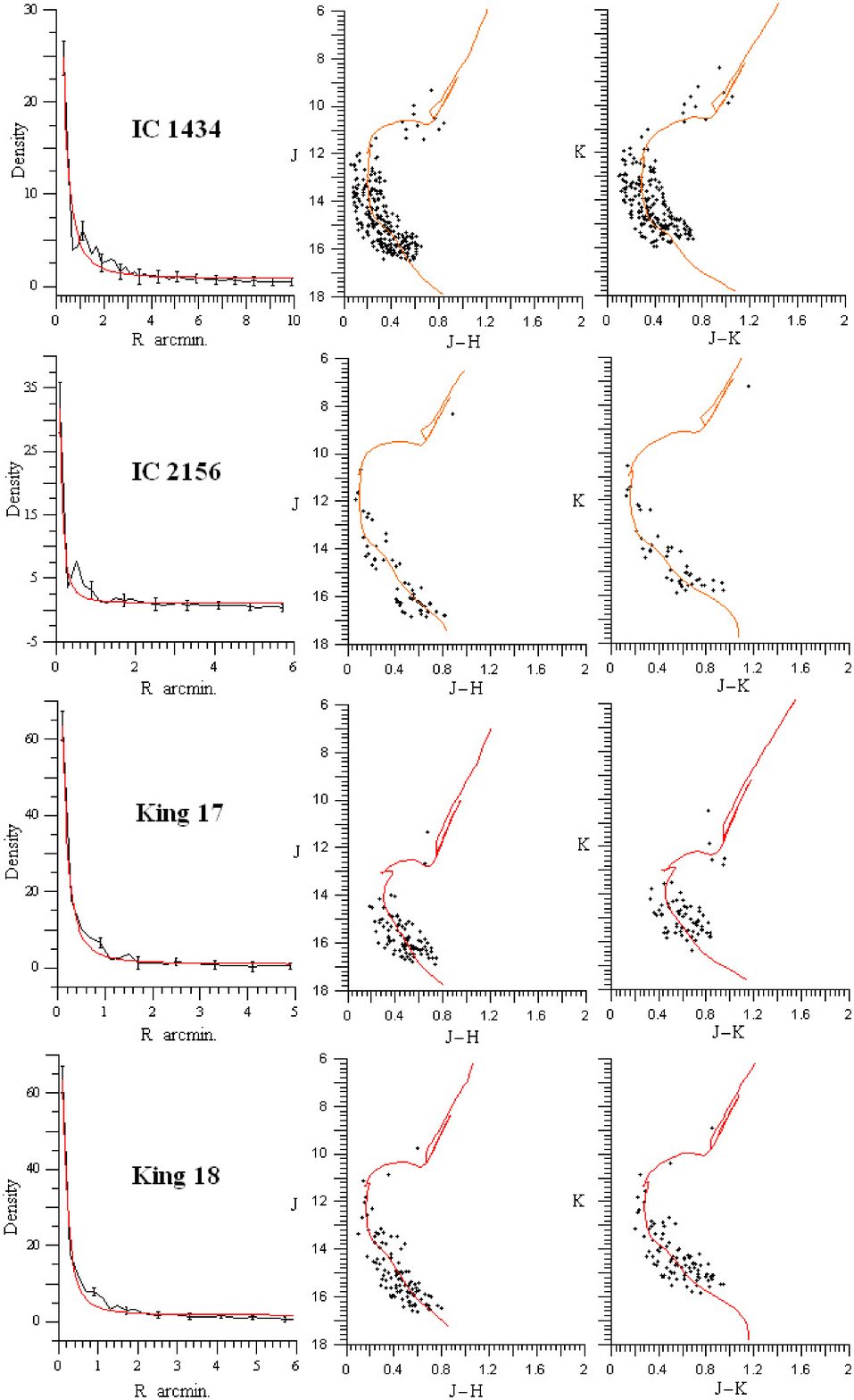}}
      \end{center}
      \caption{The radial stellar density distribution, and CMDs with isochrone fits for
 the investigated clusters from "\astrobj{IC 1434}" to "\astrobj{King 18}".}
      \end{figure*}

   \begin{figure*}
   \begin{center}
      {\includegraphics[width=15cm,height=22cm]{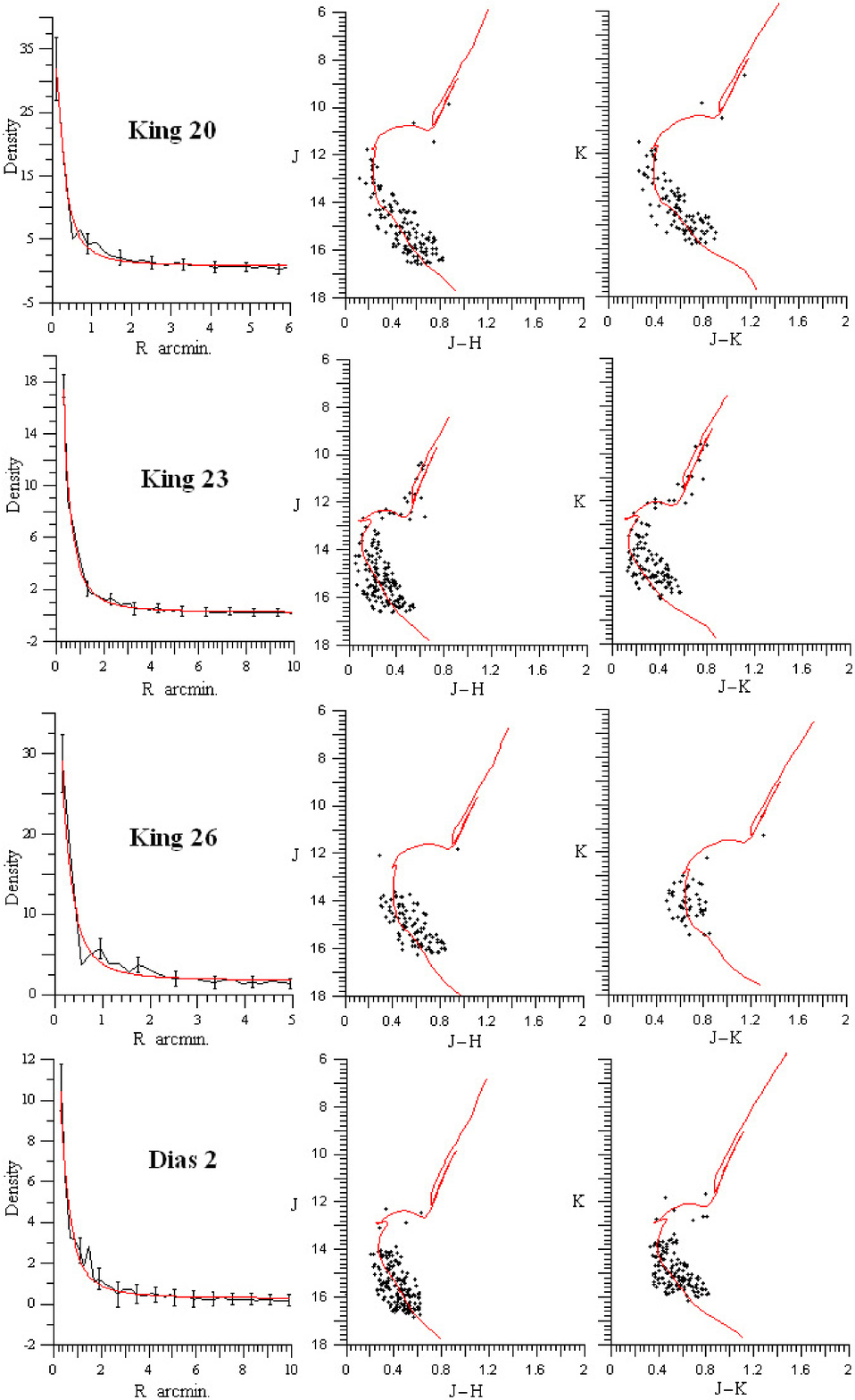}}
\end{center}
      \caption{The radial stellar density distribution, and CMDs with isochrone fits for
 the investigated clusters from "\astrobj{King 20}" to "\astrobj{Dias 2}".}
      \end{figure*}

\begin{figure*}
   \begin{center}
      {\includegraphics[width=15cm,height=22cm]{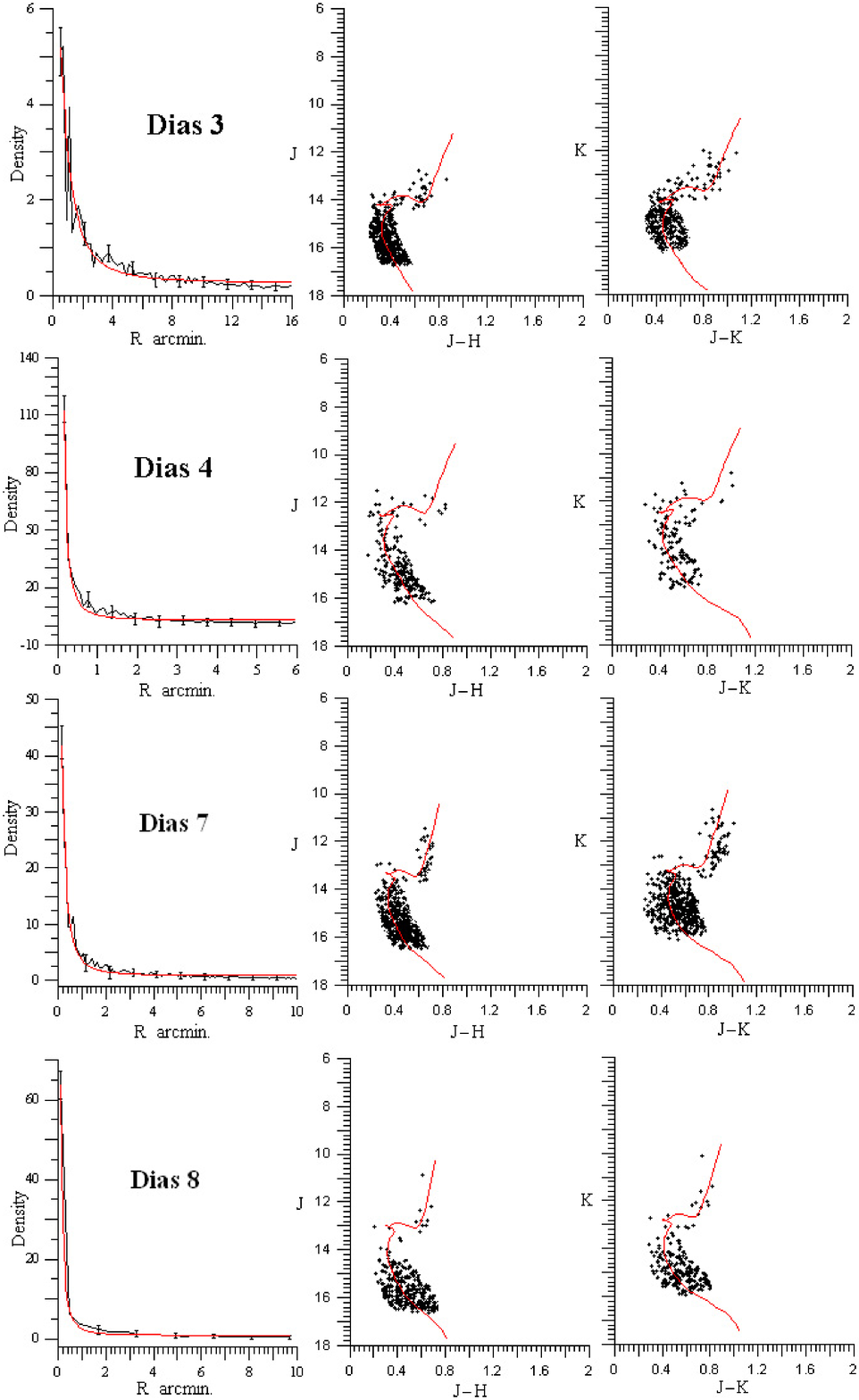}}
\end{center}
      \caption{The radial stellar density distribution, and CMDs with isochrone fits for
 the investigated clusters from "\astrobj{Dias 3}" to "\astrobj{Dias 8}".}
       \end{figure*}


\begin{table*}
\renewcommand{\arraystretch}{1}
\caption[]{The equatorial and galactic coordinates and the diameters of the investigated clusters, as taken
from "Webda" and "Dias" catalogs.}\label{tab1}
\begin{minipage}{\textwidth}
\begin{tabular}{lccrcc} \hline

            Cluster
            & $\alpha~^{h}~^{m}~^{s}$
            & $\delta~^{\circ}~{'}~{''}$
            & G. Long.$^{\circ}$
            & G. Lat.$^{\circ}$
            & Diameter~$^{'}$\\
\hline

\astrobj{IC 1434}   &  22:10:30  &  +52:50:00  & 99.937   & --2.700 &  6  \\
\astrobj{IC 2156}   &  06:04:51  &  +24:10:00  & 186.291  & +1.297  &  3  \\
          &            &             &          &         &     \\
\astrobj{King 17}   &  05:08:24  &  +39:05:00  & 167.291  & --0.731 &  5  \\
\astrobj{King 18}   &  22:52:06  &  +58:17:00  & 107.768  & --1.031 &  5 \\
\astrobj{King 20}   &  23:33:18  &  +58:29:00  & 112.853  & --2.851 &  5  \\
\astrobj{King 23}   &  07:21:48  &  --00:59:00 & 215.529  & +7.202  &  5  \\
\astrobj{King 26}   &  19:29:00  &  +14:52:00  & 50.409    & --1.339 &  2 \\
          &            &             &          &         &    \\
\astrobj{Dias 2}    &  06:09:09  &  +04:35:24  & 203.965  & --7.254 &  5 \\
\astrobj{Dias 3}    &  07:10:28  &  --08:26:14 & 222.603  & --0.350 &  12 \\
\astrobj{Dias 4}    &  13:43:40  &  --63:01:30 & 308.814  & --0.751 &  1.4 \\
\astrobj{Dias 7}    &  19:49:22  &  +21:09:48  & 58.276   & --2.461 &  1.7 \\
\astrobj{Dias 8}    &  19:52:07  &  +11:37:54  & 50.335   & --7.827 &  2.3 \\
\hline
\end{tabular}
\end{minipage}
\end{table*}

\begin{table*}
\renewcommand{\arraystretch}{1.2}
\renewcommand{\tabcolsep}{1.6 mm}
\caption[]{The redetermined coordinates and
diameters of the investigated clusters with the derived
astrophysical main parameters. Columns display, respectively,
cluster name, center coordinates, angular diameter, age,
reddening, distance modulus, distance from the sun, distance from
the galactic center, the projected distances on the galactic plane
from the sun, and the distance from galactic plane.}\label{tab3}
\centering
\begin{minipage}{\textwidth}
\begin{tabular}{lccccrcrrrrcc}

\hline
            Cluster
            & $\alpha$
            & $\delta$
            & Diameter
            & Age
            & E$_{B-V}$
            & m-M
            & Distance\,    \,
            & $R_\mathrm{gc}$
            & $X_{\odot}$
            & $Y_{\odot}$
            & $Z_{\odot}$   \\
& $~~^{h}~~^{m}~~^{s}$ & $~~^{\circ}~~{'}~~{''}$ & {\it arcmin.} & {\it Gyr} & {\it mag\,\,} & {\it mag} & {\it pc\,  \,  \,     \,} & {\it kpc} & {\it pc} & {\it pc} & {\it pc} \\
\hline

\astrobj{IC 1434}   & 22:10:34 & +52:49:40   & 7.0  & 0.32 & 0.66 & 13.0 & 3035 $\pm$ 140 & 9.5  & 523    & 2986   & --143 \\
\astrobj{IC 2156}   & 06:04:51 & +24:09:30   & 4.0  & 0.25 & 0.67 & 12.2 & 2100 $\pm$ ~95 & 10.6 & 2087   & --230  & 47  \\
          &          &             &      &      &      &      &                &      &        &        &       \\
\astrobj{King 17}   & 05:08:25 & +39:05:04   & 5.6  & 0.79 & 0.73 & 13.0 & 2960 $\pm$ 135 & 11.4 & 2887   & 650    & --38 \\
\astrobj{King 18}   & 22:52:07 & +58:16:57   & 4.8  & 0.35 & 0.52 & 11.8 & 1860 $\pm$ ~85 & 9.2  & 567    & 1770   & --33 \\
\astrobj{King 20}$^{*}$   & 23:33:18 & +58:27:52   & 6.0  & 0.28 & 0.67 & 12.2 & 2100 $\pm$ ~95 & 9.5  & 815    & 1930   & --104 \\
\astrobj{King 23}   & 07:21:47 & --00:59:06  & 7.2  & 0.89 & 0.16 & 12.6 & 3113 $\pm$ 140 & 11.2 & 2513   & --1795 & 390 \\
\astrobj{King 26}   & 19:59:01 & +14:52:02   & 4.4  & 0.44 & 1.27 & 13.2 & 2600 $\pm$ 120 & 7.1  & --1656 & 2003   & --61 \\
          &          &             &      &      &      &      &                &      &        &        &       \\
\astrobj{Dias 2}    & 06:09:11 & +04:35:35   & 11.0 & 0.79 & 0.61 & 12.8 & 2835 $\pm$ 130 & 11.2 & 2570   & --1142 & --358 \\
\astrobj{Dias 3}    & 07:10:31 & --08:25:39  & 16.0 & 1.41 & 0.64 & 13.9 & 4650 $\pm$ 215 & 12.3 & 3423   & --3147 & 28 \\
\astrobj{Dias 4}    & 13:43:25 & --63:00:48  & 6.4  & 1.26 & 0.60 & 12.2 & 2150 $\pm$ 100 & 7.3  & --1347 & --1675 & --28 \\
\astrobj{Dias 7}    & 19:49:21 & +21:10:12   & 10.0 & 2.00 & 0.42 & 12.4 & 2540 $\pm$ 115 & 7.5  & --1334 & 2158   & --109 \\
\astrobj{Dias 8}    & 19:52:06 & +11:38:04   & 10.0 & 2.24 & 0.30 & 12.0 & 2220 $\pm$ 100 & 7.3  & --1404 & 1693   & --302 \\

\hline
\end{tabular}
\end{minipage}
\begin{list}{}
\item $^{*}$ {{\it From Bica et al. (2006)}: Diam.=6.0 {\it arcmin.}; Age=0.2 {\it Gyr}; E$_{B-V}$=0.65 {\it mag}; Distance=1900 {\it pc}}.
\end{list}
\end{table*}

\end{document}